\documentclass[11pt,twoside]{article}

\usepackage{asp2006}
\usepackage{epsfig,color}

\def\spose#1{\hbox to 0pt{#1\hss}}
\def\ltsimm{\mathrel{\spose{\lower 3pt\hbox{$\sim$}}
        \raise 2.0pt\hbox{$<$}}}
\def\gtsimm{\mathrel{\spose{\lower 3pt\hbox{$\sim$}}
        \raise 2.0pt\hbox{$>$}}}

\def\km{{\rm\thinspace km}}
\def\cm{{\rm\thinspace cm}}

\def\s{{\rm\thinspace s}}
\def\yr{{\rm\thinspace yr}}
\def\g{{\rm\thinspace g}}
\def\kmps{\hbox{${\rm\km\s^{-1}\,}$}}

\def\erg{{\rm\thinspace erg}}

\def\Hz{{\rm\thinspace Hz}}

\def\ster{{\rm\thinspace ster}}
\def\ergps{\hbox{${\rm\erg\s^{-1}\,}$}}
\def\Rsol{\hbox{${\rm\thinspace R_{\odot}}$}}
\def\Msol{\hbox{${\rm\thinspace M_{\odot}}$}}

\def\Msolpyr{\hbox{${\rm\Msol\yr^{-1}\,}$}}
\def\pcm{\hbox{${\rm\cm^{-1}\,}$}}
\def\pcm2{\hbox{${\rm\cm^{-2}\,}$}}
\def\pcm3{\hbox{${\rm\cm^{-3}\,}$}}
\def\ergpscm3Hz{\hbox{${\rm\ergps\cm^{-3}\Hz^{-1}\,}$}}
\def\ergpscm3Hzster{\hbox{${\rm\ergps\cm^{-3}\Hz^{-1}\ster^{-1}\,}$}}
\def\gpcm3{\hbox{${\rm\g\cm^{-3}\,}$}}
\def\ergpcm2{\hbox{${\rm\erg\cm^{-2}\,}$}}
\def\ergpcm3{\hbox{${\rm\erg\cm^{-3}\,}$}}
\def\phpscm2{\hbox{${\rm photons\s^{-1}\cm^{-2}\,}$}}

\begin{document}
\title{Models of the non-thermal emission from early-type binaries}
\author{J. M. Pittard}
\affil{The University of Leeds, Leeds, LS2 9JT, U.K.}

\begin{abstract}
The powerful wind-wind collision in massive star binaries creates a
region of high temperature plasma and accelerates particles to
relativistic energies. I briefly summarize the hydrodynamics of the
wind-wind interaction and the observational evidence, including recent
$\gamma$-ray detections, of non-thermal emission from such systems. I
then discuss existing models of the non-thermal emission and their
application to date, before concluding with some future prospects.

\end{abstract}

\section{Introduction}
The winds of O and WR stars couple high mass-loss rates of $\sim
10^{-7}$ to a few times $10^{-5}\,\Msolpyr$ with fast velocities
typically $\sim 1000-3000\,\kmps$. The ram pressure balance between
the hypersonic winds determines the position of the wind-wind
collision region (WCR). Shocks are formed either side of the WCR which
thermalize the plasma, heating it to $\sim 10^{7}-10^{8}\,$K. The
properties (e.g. density) of the WCR can span a very large range
\citep[see e.g. Table~2 in][]{Pittard:2005}, reflecting the diversity
of the underlying binary population. In systems where the orbital
period is only a few days, the shocks are collisional, and the WCR
displays a large aberration and downstream curvature due to the
coriolis force. Cooling of the shocked gas is also likely to be
significant.  In addition, the presence of the companion star's
radiation field may have significant influence on the driving and
dynamics of the winds.  In contrast, in systems with longer orbital
periods the shocks may be collisionless, orbital and
radiation-field-induced effects on the dynamics and geometry of the
WCR are much smaller, and the hot shocked gas may only cool
adiabatically as it flows downstream. Particle acceleration at the
shocks (but also possibly within the WCR - see \citet{Pittard:2006b}
for a detailed review) is likely to occur in these wider
systems. Systems with (highly) eccentric orbits are particularly
interesting since the changing separation of the stars is useful as a
probe of the physics which takes place. Colliding wind binaries (CWBs)
are also useful as simpler, less complicated, analogues of the
multiple wind-wind collisions which occur throughout the volume of
clusters of massive stars.

The collision of the winds is best studied using hydrodynamical
simulations. A seminal paper by \citet*{Stevens:1992} was the first to
self-consistently include cooling, and it focussed on the dynamical
instabilities and X-ray emission which arise from the WCR.  Other
works since then have examined the influence of the companion star's
radiation field \citep{Stevens:1994,Gayley:1997}, the effects of
thermal conduction \citep{Myasnikov:1998}, electron-ion temperature
non-equilibrium \citep{Zhekov:2000}, and non-equilibrium ionization
\citep{Zhekov:2007}. The effect of clumps within the stellar winds on
a WCR near the radiative and adiabatic limits has been investigated by
\citet{Walder:1998} and \citet{Pittard:2007}, respectively.
Three-dimensional models have allowed orbital effects to be explored
\citep*{Lemaster:2007}. This work has also been complemented by
smoothed-particle-hydrodynamics simulations \citep{Okazaki:2008} and
the development of a compute-efficient dynamical model
\citep{Parkin:2008,Parkin:2009}.  More recently, three dimensional
hydrodynamical models which incorporate gravity, the driving of the
winds, the cooling of the shocked plasma, and the orbital motion of
the stars, have been presented by \citet{Pittard:2009}.  This work
represents the most realistic simulation to date of the wind-wind
collision in short period systems.

\begin{figure*}
\begin{center}
\psfig{figure=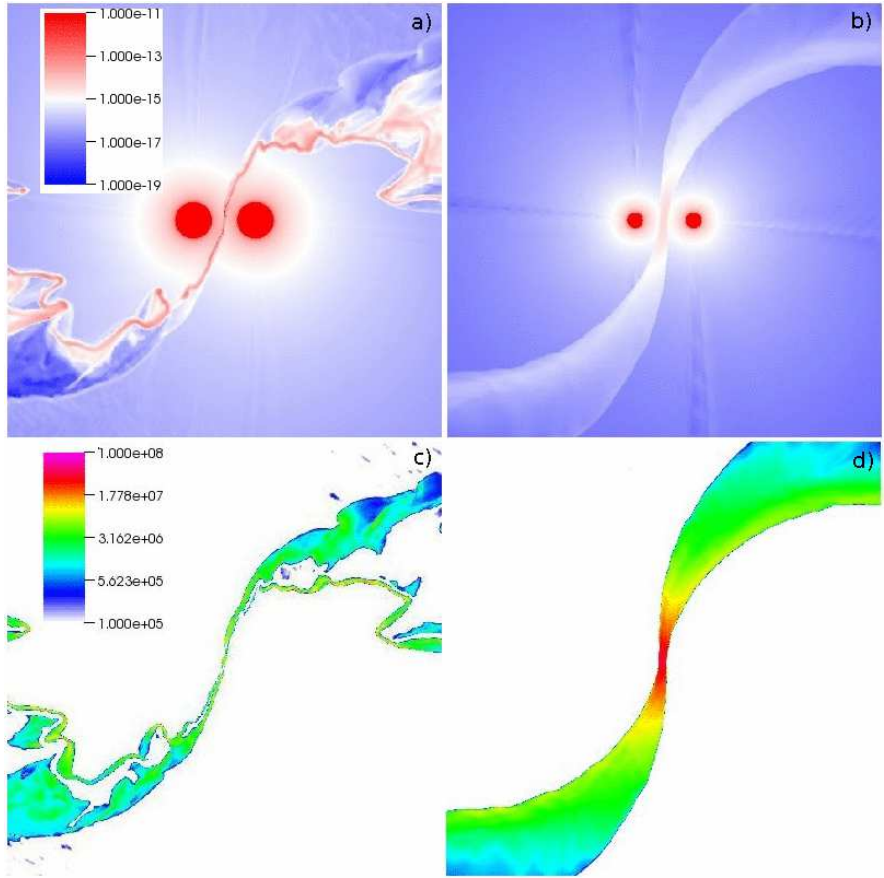,width=11.7cm}
\end{center}
\caption[]{Density and temperature plots through the orbital plane of
hydrodynamic models of the wind-wind collision in short period O6V+O6V
binaries. In the left plots the orbital period is 3 days, and the
winds collide at relatively low speeds. Hence, the WCR is highly
radiative and is strongly distorted by the orbital motion of the
stars.  In the right plots the orbital period is 10 days, the winds
collide at higher speeds, and the WCR is largely adiabatic.  The top
panels display the density and the bottom panels the temperature.  The
left panels have sides of length $240\;\Rsol$, while the right panels
have sides of length $570\;\Rsol$. See \citet{Pittard:2009} for
further details.}
\label{fig:3d_ooraddrv}
\end{figure*}

Fig.~\ref{fig:3d_ooraddrv} displays cuts through the orbital plane of
the density and temperature in two different systems with identical
O-stars but with orbital periods of 3 (left) and 10 days
(right). The WCR is highly radiative in the shorter period system
since the winds collide before they have had much opportunity to
accelerate, the postshock gas temperature is lower, and the postshock
density is higher. There is also considerable inhibition of the acceleration
of the winds caused by the opposing radiation field of the
companion star.  In contrast, with an orbital period of 10 days the
wider separation of the stars is enough to more than double the pre-shock
velocities of the winds (along the line-of-centres through the stars
it increases from $730\,\kmps$ to $1630\kmps$), which leads to
significantly higher (lower) postshock temperatures (densities),
resulting in a largely adiabatic WCR. The higher wind speeds and
reduced orbital velocities also diminishes the aberration and
downstream curvature of the WCR due to coriolis forces (note the
difference in spatial scale between the two systems presented in
Fig.~\ref{fig:3d_ooraddrv}).

\section{Particle Acceleration in Colliding Wind Binaries}
Direct evidence for non-thermal emission from relativistic particles
within a WCR was presented by \citet{Williams:1997}. Here, radio
images were overlaid on UKIRT shift-and-add IR images of WR~147. When
the southern (thermal) radio source was aligned with the southern (WR)
star, the northern (non-thermal) radio source was found to lie just
south of the northern (O) star, in a position consistent with the
point of ram-pressure balance between the winds\footnote{WR\,147 is 
the only CWB to date which also has spatially
resolved X-ray emission - see \citet{Pittard:2002}.}. Direct imaging of the
WCR in WR~146 and WR~140 have provided further support for this
interpretation \citep{Dougherty:2000,Dougherty:2005}. See Dougherty
(these proceedings) for further details.

WR~146 and WR~147 are both very wide, and the non-thermal radio
emission from the WCR escapes easily from the system. This contrasts
with WR~140 (the best studied of any WR+O binary), where the stars are
very much closer together, and which displays dramatic,
phase-repeatable, variations in its radio emission over the course of
every 7.94~yr orbit \citep[see][]{White:1995,Dougherty:2005}.  The
high eccentricity of the orbit causes the stellar separation to vary
between $\approx 2-28$\,AU. While the radio lightcurve has defied
satisfactory explanation to date, it is likely that at least part of
its variation is caused by variable circumstellar extinction to the
source of the non-thermal emission as the O star orbits in and out of
the radio photosphere in the dense WR wind. More recently, this system
has been imaged with the VLBA, yielding a full orbit definition,
including, most importantly, the inclination of the system
\citep{Dougherty:2005}.  While this study has helped to provide some
of the best modelling constraints of any system, relatively little is
known about the wind of the O star, and the wind momentum ratio of the
system remains ill-constrained.

In the hard X-ray and $\gamma$-ray regimes there has been many false
dawns concerning potential emission from CWBs. However, definitive
evidence for non-thermal X-ray and $\gamma$-ray emission from CWBs has
finally been presented in recent years, with an especially exciting
period in just the last few months. Using INTEGRAL,
\citet*{Leyder:2008} presented a clear detection of MeV $\gamma$-ray
emission from $\eta$\,Carinae \citep[a lower spatial resolution
observation with BeppoSAX was previously presented by][though some of
the emission attributed to $\eta$\,Carinae was shown by
\citet{Leyder:2008} to be associated with other nearby
sources]{Viotti:2004}. This detection has since been confirmed with a
Suzaku observation \citep{Sekiguchi:2009}.  And in just the last few
months, $\eta$\,Carinae has been associated with an AGILE source
\citep{Tavani:2009}, and is on the FERMI bright source list.
Upper limits have also been placed on the MeV emission from WR\,140,
WR\,146, and WR\,147 \citep{DeBecker:2007}, and on the TeV emission
from WR\,146 and WR\,147 \citep{Aliu:2008}. Finally, a great
deal of effort has gone into searching for non-thermal X-ray emission
from CWBs with known non-thermal radio emission 
\citep[e.g.][]{Rauw:2002,DeBecker:2004,DeBecker:2006}. However,
the lack of success in this undertaking appears to show 
that the physical conditions necessary to produce
{\em observable} emission at radio and hard X-ray energies are
mutually incompatible.

There are many possible mechanisms for accelerating particles in
CWBs. Most works assume that diffusive shock acceleration (DSA)
is the dominant process, though reconnection and turbulent processes
are other possibilities. Each mechanism differs in its efficiency,
and in the properties of the non-thermal particles which it produces,
such as their spectral index. A detailed discussion of the many
possibilities can be found in \citet{Pittard:2006b}.
 
\section{Models of the non-thermal emission}
\subsection{Early models}
Some notes on the key physics of particle acceleration in CWBs
were presented by \citet{Eichler:1993}.  Early models of the
non-thermal radio emission were very simple, with the observed flux
($S_{\nu}^{\rm obs}$) assumed to be a combination of the free-free
flux from the spherically symmetric winds ($S_{\nu}^{\rm ff}$), plus
the flux from a point-like non-thermal source located at the
stagnation point of the winds ($S_{\nu}^{\rm nt}$), the latter being
attenuated by free-free absorption (opacity $\tau_{\nu}^{\rm ff}$)
through the surrounding winds:
\begin{equation}
\label{eq:simplerad}
S_{\nu}^{\rm obs} = S_{\nu}^{\rm ff} + S_{\nu}^{\rm nt}\,{\rm e}^{-\tau_{\nu}^{\rm ff}}.
\end{equation}
This approach allows relatively simple analytical solutions to the
radiative transfer equation to be obtained
\citep[e.g.][]{Williams:1990,Chapman:1999}. Unsurprisingly, such
simple models fail to reproduce the spectral variation of the emission
with orbital phase. This led \citet{Williams:1990} to propose the need
for a more complicated model which accounted for the low-opacity
``hole'' in the dense WR wind created by the O-star's wind. However,
\citet{White:1995} pointed out that in the case of WR\,140, even the
O-star's wind has significant opacity. These works clearly demonstrate
the need for more realistic modelling where the presence of the
hot, low-opacity, WCR is a fundamental component. In addition, the
interferometric observations of spatially-extended synchrotron
emission indicate that the assumption of a point-like non-thermal
source needs also to be modified. Other problems with these simple
models are that no consideration of cooling (e.g. inverse
Compton) or other mechanisms (e.g. the Razin effect) are made. 

In many ways, the state of modelling in the $\gamma$-ray domain was, until
a few years ago, even more rudimentary. With no firm detections at
$\gamma$-ray energies, theoreticians were reduced to estimating the 
inverse Compton (IC) luminosity,
$L_{\rm ic}$, by the following simple formula:
\begin{equation}
\label{eq:simpleic}
L_{\rm ic} = \frac{U_{\rm ph}}{U_{\rm B}} L_{\rm sync},
\end{equation}
where $L_{\rm sync}$ is the synchrotron luminosity, and $U_{\rm ph}$
and $U_{\rm B}$ are the photon and magnetic field energy densities,
respectively. A fundamental problem with the use of the above formula
is that the predicted value of $L_{\rm ic}$ is highly sensitive to the
assumed magnetic field ($B = \sqrt{8 \pi U_{\rm B}}$). Varying $B$
results in a wide range of predictions for $L_{\rm ic}$, as shown in
\citet{Benaglia:2003}. The magnetic field in the WCR is highly
uncertain, even when the surface magnetic fields are known (which is
very rare), because one cannot necessarily extrapolate these to obtain
their strength in the WCR, since there are the possibilities of
large-scale magnetic reconnection within the WCR on the one hand, and
magnetic field amplification by non-linear DSA on the other.  Finally,
{\em observed} rather than {\em intrinsic} values have been used for
$L_{\rm sync}$. This may be of little consequence in the wider systems
where the attenuation of non-thermal emission through the
circumstellar envelope surrounding the stars may be negligible, but
for a given set of parameters, $L_{\rm ic}$ could be underestimated in
closer systems.

While there are many problems and uncertainties associated with the
early modelling work of the non-thermal radio and $\gamma$-ray
emission from CWBs, these simple models nevertheless paved the way for
the more sophisticated modelling which has since followed, as we now
describe.

\subsection{Recent developments}
\subsubsection{Models of the radio emission}
A major step along the path towards improved models of the radio
emission from colliding wind binaries was taken by
\citet{Dougherty:2003}. This work removed the assumptions of a
point-like source of non-thermal emission, and a spherically
symmetric, single temperature, surrounding envelope. Instead, models
of the thermal and non-thermal radio emission were based on 2D,
axisymmetric hydrodynamical simulations. This approach provided a much
better description of the density and temperature distribution in the
the system, allowing sight-lines to the observer to pass through
regions of both high and low opacity. The non-thermal emission
was treated in a phenomenological way: accelerated electrons
were assumed to be present within the WCR, with an energy density 
($U_{\rm rel,e}$) proportional to the local thermal energy density
($U_{\rm th}$) i.e. $U_{\rm rel,e} = \zeta_{\rm rel,e} U_{\rm th}$.
The magnetic field energy density was specified in a similar manner:
$U_{\rm B} = \zeta_{\rm B} U_{\rm th}$. The non-thermal electrons
were further assumed to have a power-law distribution,
$N(\gamma)d\gamma = C \gamma^{-p}d\gamma$, where $\gamma$ is the
Lorentz factor and $C$ is proportional to 
$\zeta_{\rm rel,e}$. The non-thermal electrons were assumed to
arise from test particle DSA, where $p=2$ for strong shocks of
adiabatic index $5/3$.

Though this work introduced a number of assumptions and involved some
essentially free parameters, it nevertheless provided a great deal of
new insight into the phenomenon of radio emission from colliding
wind binaries. An immediate benefit was the realization of the
potential importance of the Razin effect in attenuating the low
frequency synchrotron emission within the WCR. It also established
several key scaling relations. For instance, given the assumptions in
the model, the total synchrotron emission from the entire WCR in
adiabatic systems was found to scale as $D^{-1/2}\nu^{-1/2}$, where
$D$ is the separation of the stars (for comparison, the X-ray emission
in the optically thin, adiabatic limit, scales as $D^{-1}$).  This
work also highlighted the importance of IC cooling, which was noted to
be important even in wide systems. Indeed, neglect of IC cooling of
the non-thermal electrons in this model led to an overestimation of
the high frequency synchrotron flux in a model of WR\,147.

\begin{figure*}
\vspace{10cm}
\includegraphics{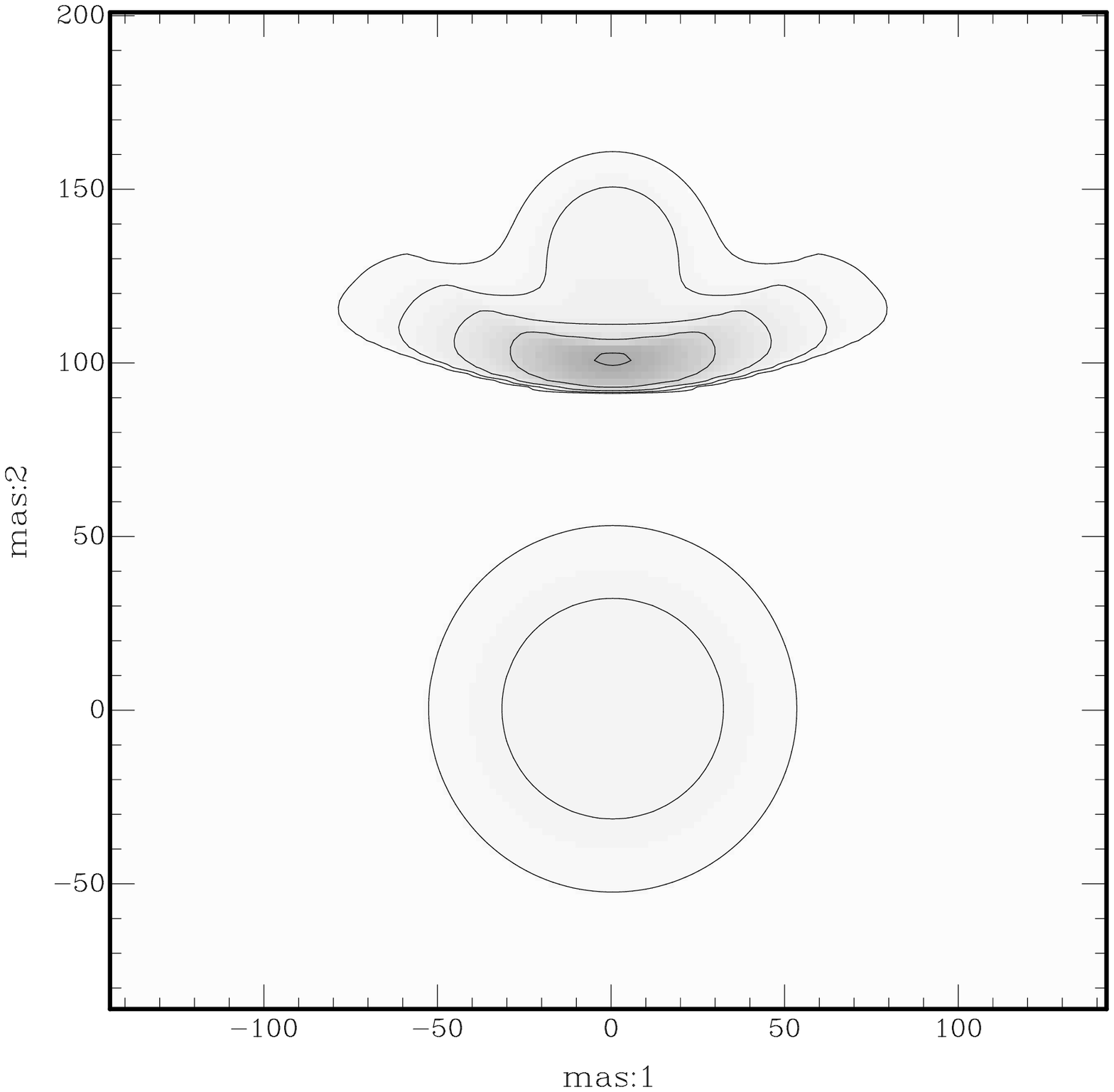}
\includegraphics{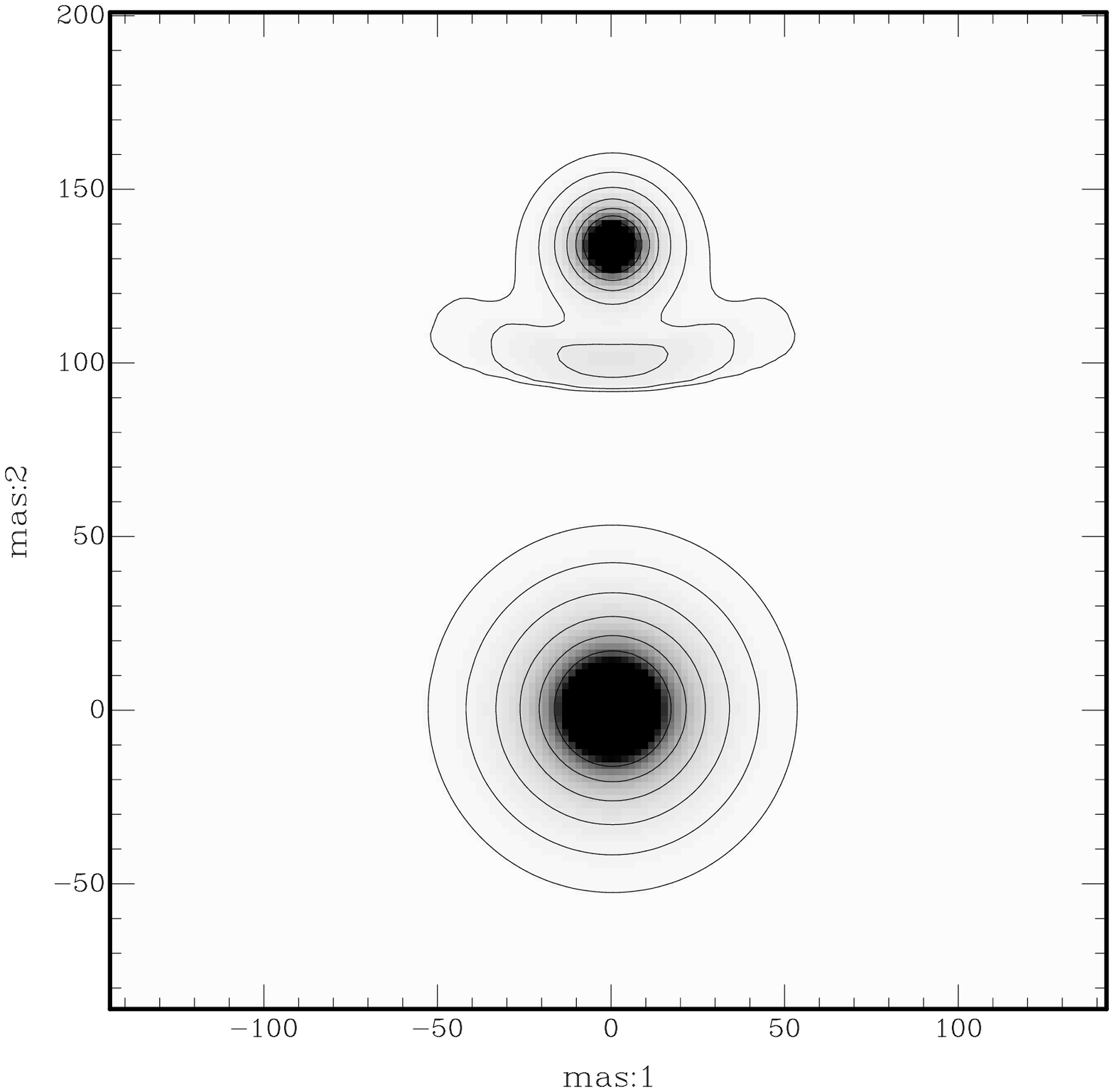}
\includegraphics{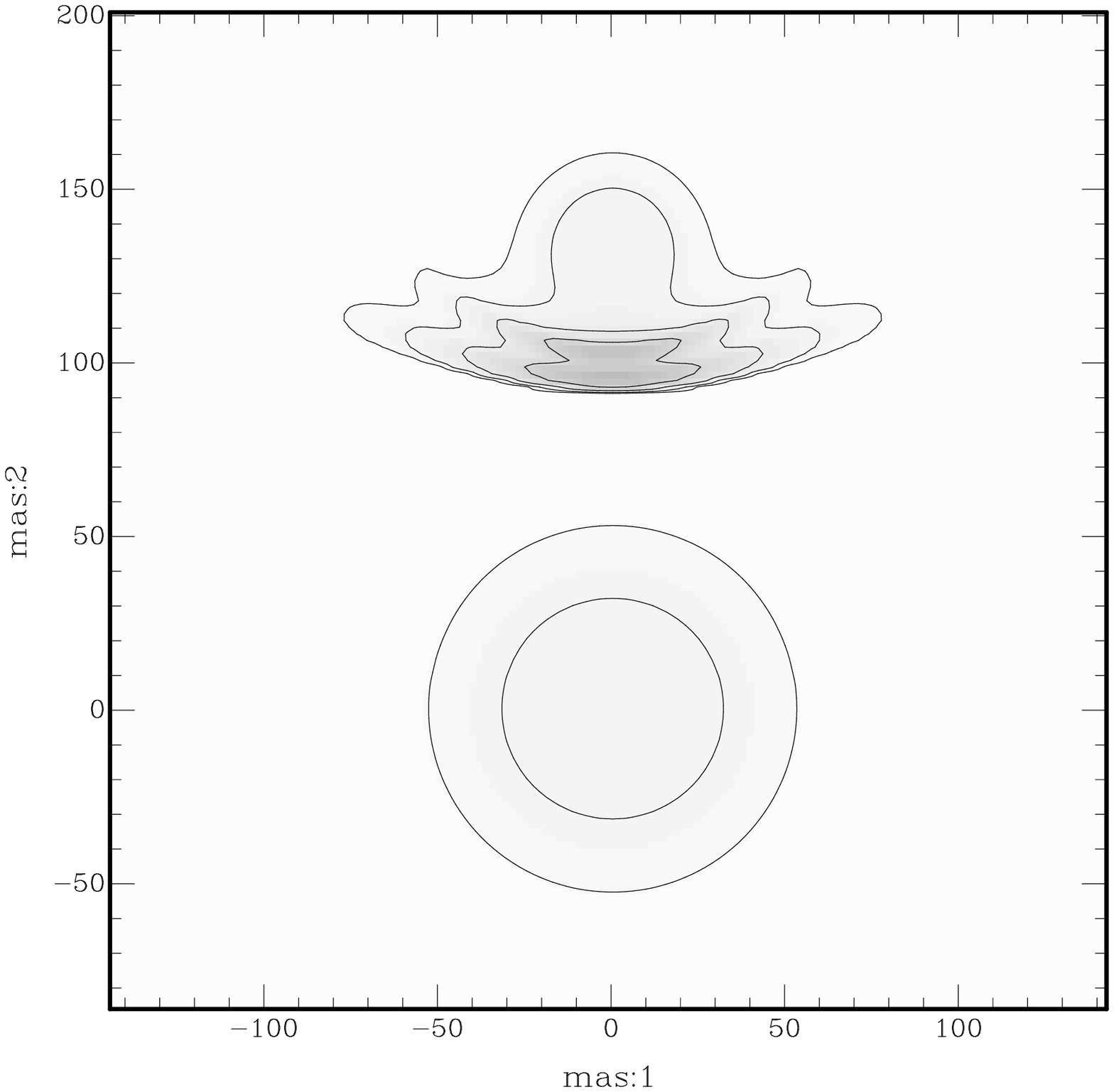}
\includegraphics{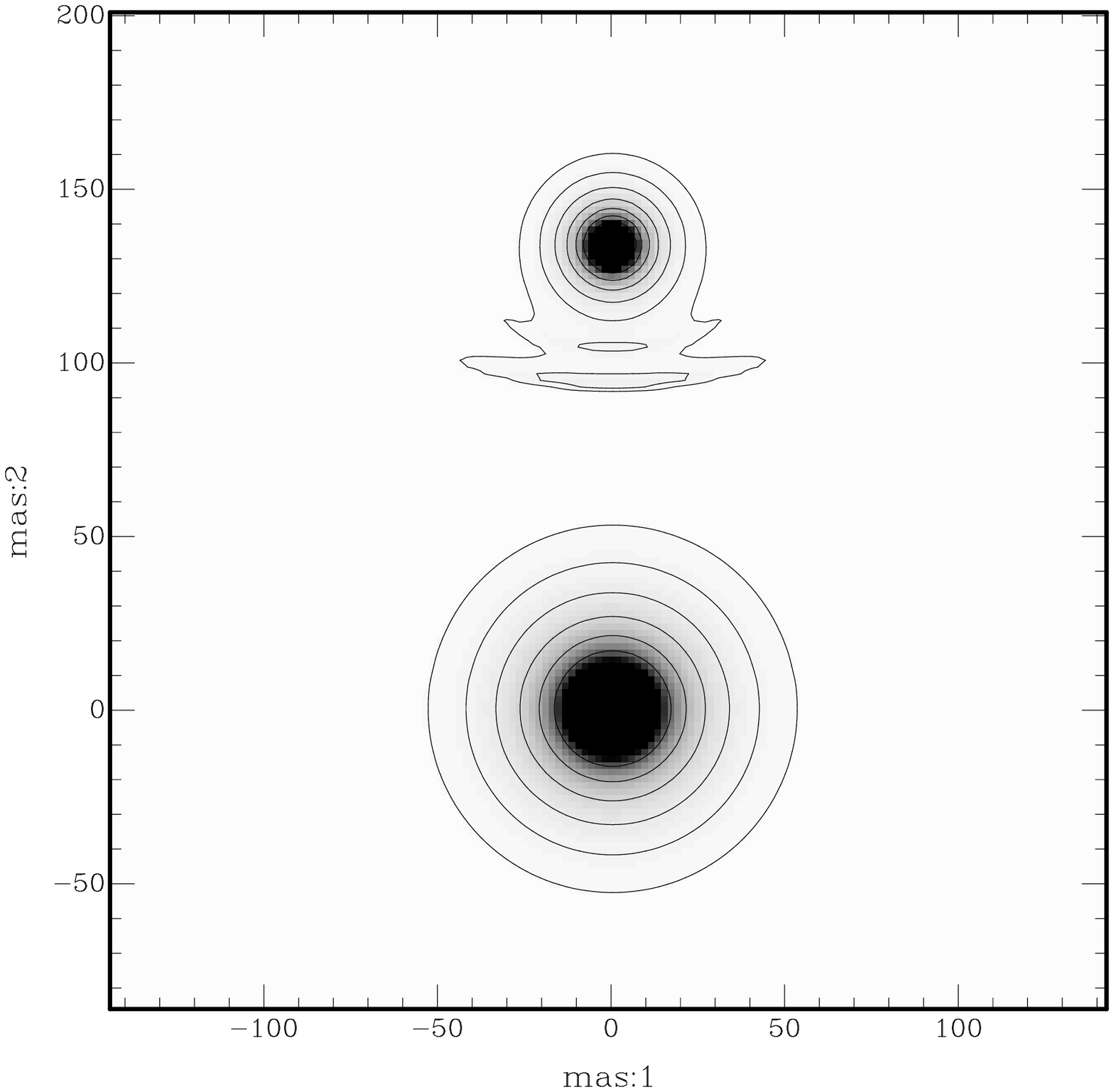}
\caption[]{The impact of IC cooling on the intensity distribution of 
  a CWB model \citep[see][for details]{Pittard:2006}, 
  for a viewing angle of $0^\circ$ and at
  1.6~GHz (top) and 22~GHz (bottom). 
  The images on the left do not include IC cooling, while
  those on the right do. Each image has the same intensity scale and
  contours.}
\label{fig:iccool}
\end{figure*}

This failing was addressed in a follow-up paper \citep{Pittard:2006}.
Here, the non-thermal electrons were assumed to be accelerated by
DSA at the {\em global} shocks confining the WCR, and to cool
once they flowed downstream. Since the amount of cooling
is dependent on the ``exposure'' time of the non-thermal electrons
to the radiation fields of the stars, this assumption led to
those electrons suffering the most severe cooling to concentrate
along the contact surface between the winds, and thus to
a dearth of emission from this region (see Fig.~\ref{fig:iccool}).
This addition significantly improved the fit between 
models  and observations of WR\,147. It also 
modified the scaling relation for the total synchrotron luminosity
noted in \citet{Dougherty:2003} - now, the {\em intrinsic} luminosity
was observed to decline with stellar separation as IC cooling
became increasingly strong. In addition, it was noted that the {\em thermal}
radio emission from the WCR scales as $D^{-1}$, in an identical
way to the thermal X-ray emission. Since this emission is
optically thin (on account of the high temperatures within the WCR),
it can mimic a synchrotron component. Therefore, one needs to
cautiously interpret data with a spectral index 
$-0.1 \ltsimm \alpha \ltsimm 0.5$, where $S_{\nu} \propto \nu^{\alpha}$
(more negative values of $\alpha$, e.g. $\alpha \approx -0.5$, 
clearly indicate a bona-fide synchrotron component).

\subsubsection{Models of the non-thermal $X$-ray and $\gamma$-ray 
emission} 
The dramatic sensitivity gains made by arrays of Cerenkov
telescopes have led in recent years to a new interest in the level of
non-thermal X-ray and $\gamma$-ray emission from colliding wind
binaries. \citet{Bednarek:2005} calculated the
expected $\gamma$-ray emission from WR\,20a, a WR+WR binary. The short orbital
period means that the optical depth to electron-positron pair creation
is high enough to initiate electromagnetic cascades in this
system. Particle acceleration by magnetic reconnection and DSA was
considered, and it was concluded that detectable neutrino fluxes
should be produced. Due to the high optical depth to
TeV photons within this system, WR\,20a cannot be directly 
responsible for nearby TeV emission \citep{Aharonian:2007}, which
more likely is the result of acceleration processes within the
collective wind of the nearby cluster Westerlund 2.

\begin{figure*}
\begin{center}
\psfig{figure=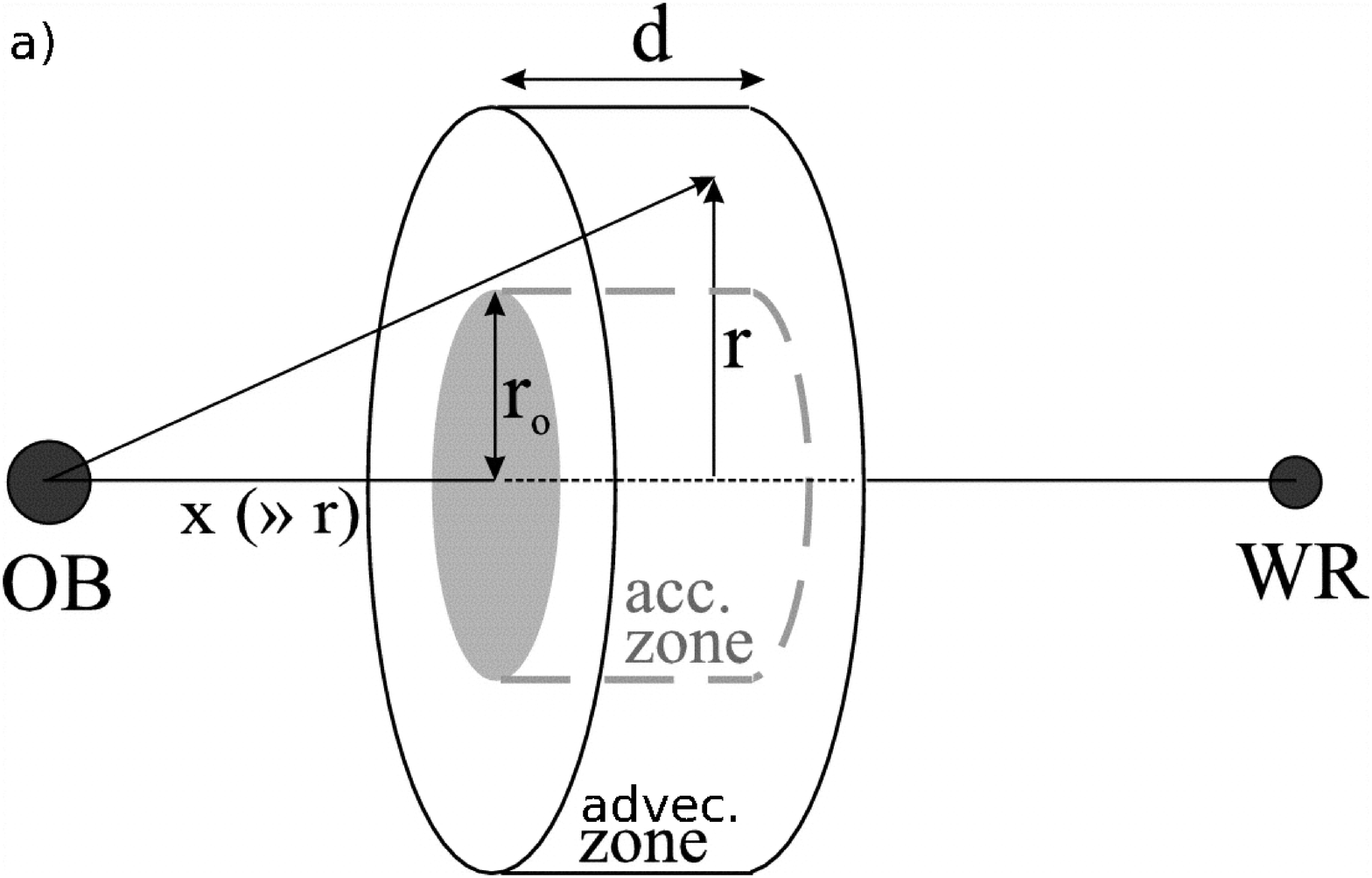,width=4.0cm}
\psfig{figure=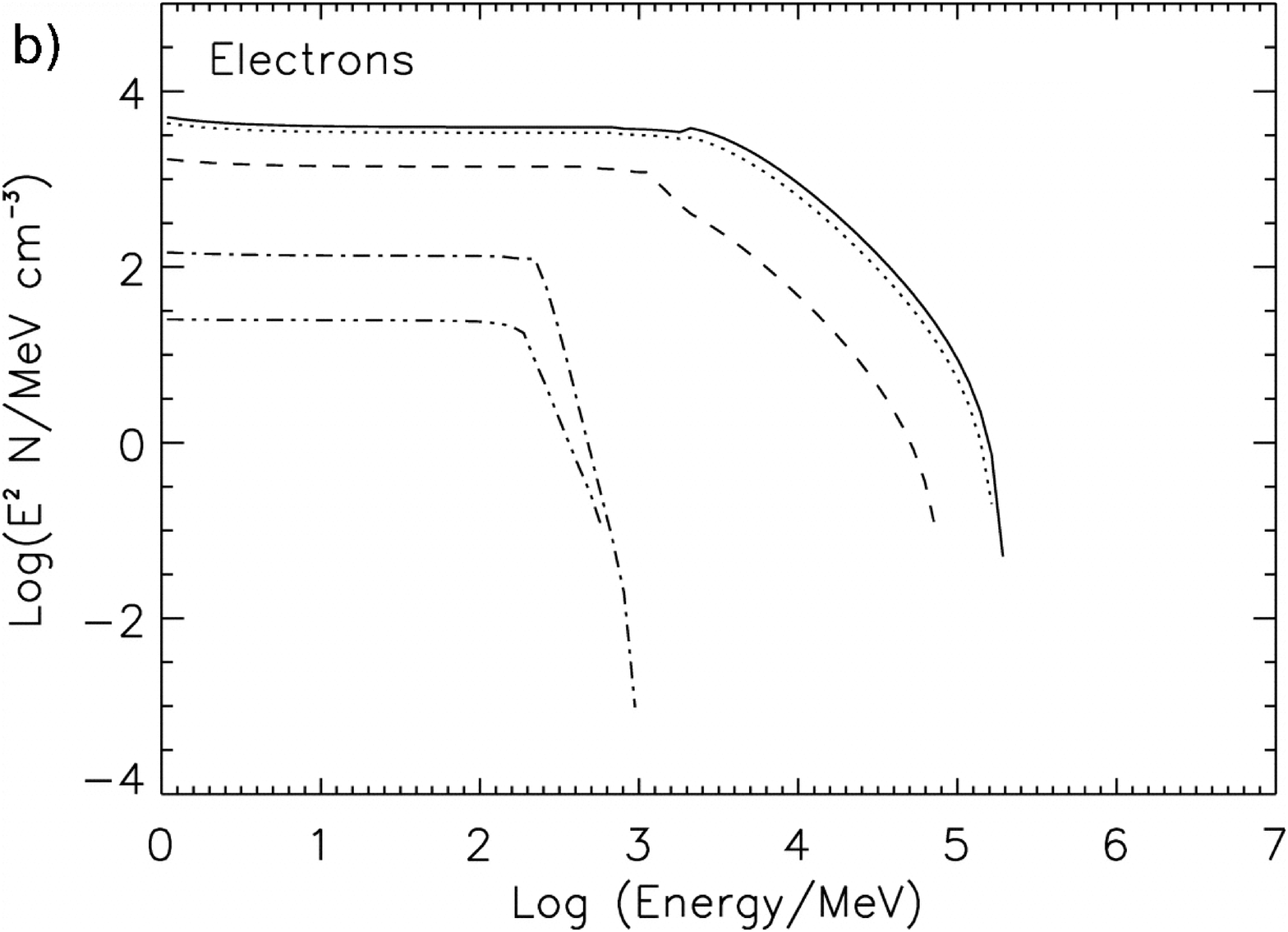,width=4.0cm}
\psfig{figure=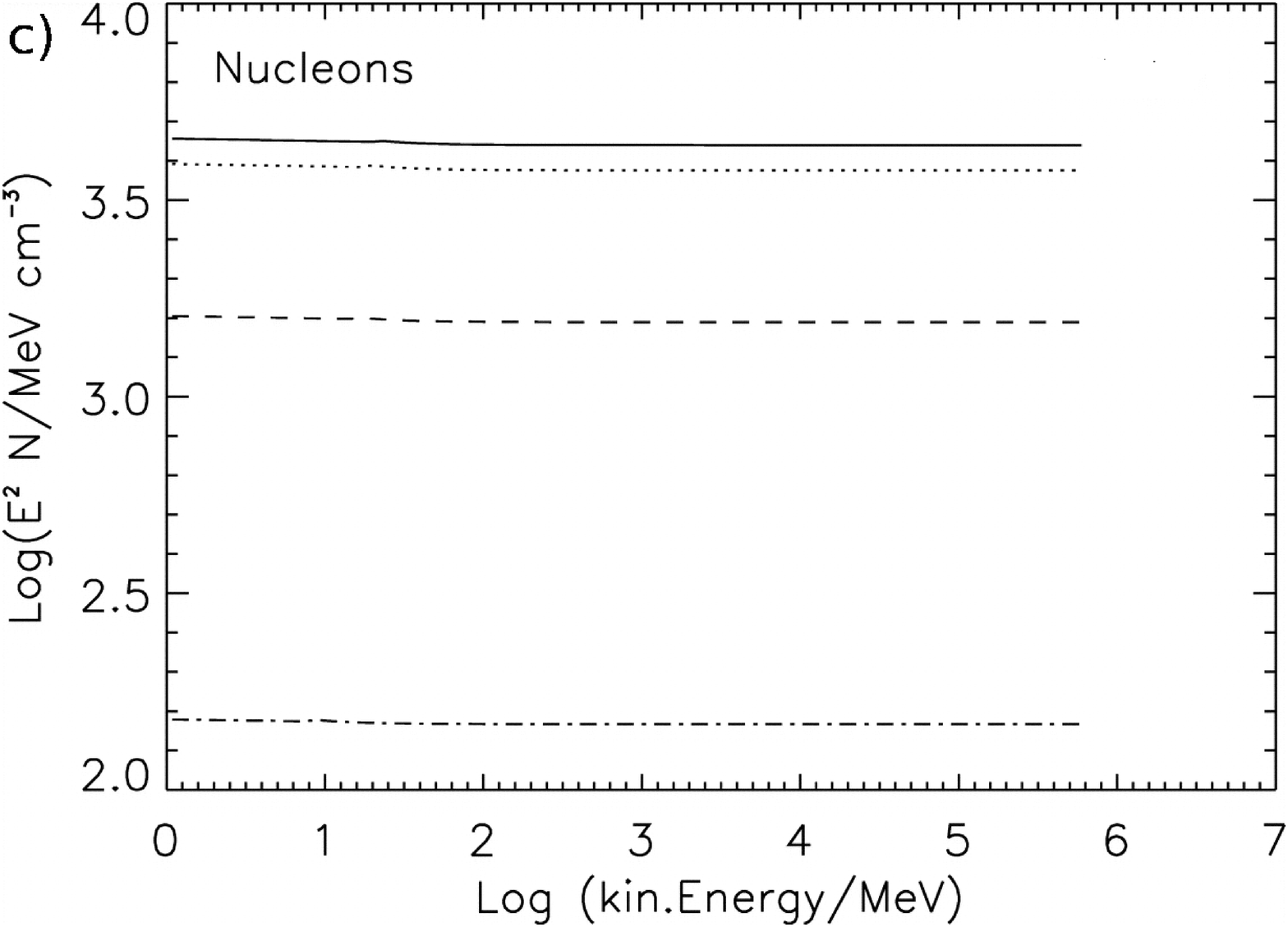,width=4.0cm}
\end{center}
\caption[]{a) Geometry of the 2-zone model in \citet{Reimer:2006}. b) Evolution
of the non-thermal electron spectrum from the inner acceleration zone (solid)
line as a function of downstream distance in the advection zone. At low 
energies adiabatic/expansion losses dominate, while at high energies IC
losses dominate. c) As b) but for nucleons. Only adiabatic losses occur.
See \citet{Reimer:2006} for further details.}
\label{fig:2zone}
\end{figure*}

Following on from this work, \citet*{Reimer:2006} developed a two-zone
model of the non-thermal emission. Particles are accelerated in 
an inner zone where their spatial diffusion exceeds their motion due to
advection with the background fluid. Their energy distribution is
self-consistently computed by 
considering all relevant gain and loss mechanisms.
Particles are assumed to be resident within this region until their
timescales for advection and diffusion are comparable,
after which they are assumed to move into the advection region where
they suffer further losses as they flow downstream. Fig.~\ref{fig:2zone}
shows the assumed geometry and the resultant non-thermal energy
spectra of the electrons and nucleons. 

\begin{figure*}
\begin{center}
\psfig{figure=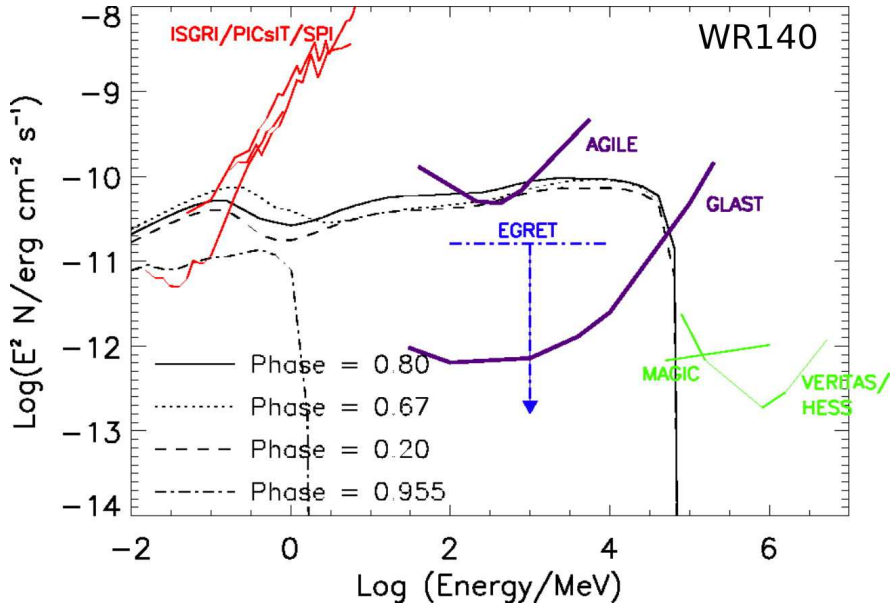,width=11.7cm}
\end{center}
\caption[]{Predicted IC spectra for WR\,140 at phases 0.2, 0.671, 0.8
and 0.955 from \citet{Reimer:2006}. $\gamma$-ray absorption is not included.}
\label{fig:reimer_wr140}
\end{figure*}

Consideration is also given to the anisotropic nature of the IC
process, where the emitted power is dependent on the scattering
angle. Fig.~\ref{fig:reimer_wr140} shows the predicted IC emission
from WR\,140 as a function of orbital phase. \citet{Reimer:2006}
conclude that while WR\,140 should be easily detected with
GLAST/Fermi, the change in the IC flux with viewing angle due to
anisotropic scattering is likely to be obscured by large variations in
the energy density of the stellar radiation fields resulting from the
high orbital eccentricity. However, the latest work by
\citet{Reimer:2009} demonstrates that it is possible to use the
property of nonisotropic IC emission to constrain the orbital
inclination of colliding wind systems.

Another model of the X-ray and $\gamma$-ray emission, which is 
in many ways complementary to that of
\citet{Reimer:2006}, was presented by \citet{Pittard:2006b}. This built
on the phenomenological model developed previously by
\citet{Dougherty:2003} and \citet{Pittard:2006} to explore the
non-thermal radio emission.  Although the energy spectrum is assumed
rather than calculated, and so in this sense is weaker than the model
in \citet{Reimer:2006}, this approach benefits from a realistic
description of the density and temperature distribution within the
system, and constraints placed on the key parameters (e.g. mass-loss
rates) by fits made to the X-ray data.  \citet{Pittard:2006b} showed
that other uncertainities, such as the particle acceleration
efficiency and the spectral index of their energy distribution (both
of which unfortunately remain ill-constrained from fits to radio
data), have at least as much influence on the predicted flux as the
angle-dependence of the IC emission.

\begin{figure*}
\begin{center}
\psfig{figure=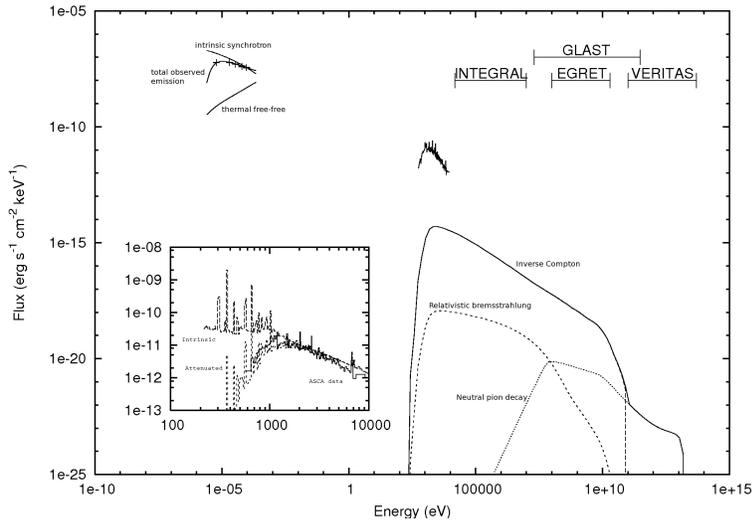,width=10.0cm}
\end{center}
\caption[]{The radio and non-thermal UV, X-ray and $\gamma$-ray
emission calculated from model~B in \citet{Pittard:2006b}, together
with the observed radio and X-ray flux (both at orbital phase
0.837). The model IC (long dash), relativistic bremsstrahlung (short
dash), and neutral pion decay (dotted) emission components are shown, along
with the total emission (solid).  See \citet{Pittard:2006b} for more
details.}
\label{fig:wr140he}
\end{figure*}

Fig.~\ref{fig:wr140he} shows a predicted spectral energy distribution
for WR\,140 from one of the models presented in \citet{Pittard:2006b}.
Large differences in the predicted $\gamma$-ray emission occur
depending on whether the low frequency turndown in the radio spectrum
results from free-free absorption through the surrounding stellar
winds, or from the Razin effect. Furthermore, satisfactory fits to the
radio spectrum at orbital phase 0.837 could be achieved in one of two
ways: either with a standard $p=2$ index, or with a harder index
(e.g. $p \approx 1.4$). Harder indices can result from the shock
re-acceleration process, whereby the non-thermal particles pass
through a sequence of shocks \citep{Pope:1994}, or from 2nd order
Fermi acceleration. Either of these processes may be significant in
CWBs, since the clumpy nature of the winds means that the WCR is
likely to be highly turbulent, with weak shocks distributed throughout
it \citep{Pittard:2007}.

While \citet{Pittard:2006b} could not determine the nature of the
absorption process from the fits to the radio spectrum, future
$\gamma$-ray detections will determine the $\gamma$-ray flux and
spectral index, and thus will also distinguish the nature of the
low-frequency turndown. The acceleration efficiency of the non-thermal
electrons and the strength of the magnetic field will then both be
revealed.

Work is also continuing on fits to the radio spectra of WR\,140
at other phases around its orbit. Preliminary results indicate that
there is significant evolution of key parameters in the model (such
as the acceleration efficiency and magnetic field).
Finally, the non-thermal emission from the short period O+O
binary models in \citet{Pittard:2009} is being investigated. One
area of interest concerns the ease with which non-thermal radio
emission can escape these systems, despite the orbital-induced 
curvature preventing lines-of-sight to the non-thermal
emission from existing purely within the low opacity, hot WCR.

\section{Conclusions}
In recent years there has been a steady improvement in theoretical
model predictions of the non-thermal emission from CWBs, in both the
radio and $\gamma$-ray domains. This work is being driven by
corresponding advances on the observational front. In the past, CWBs
have played a poor second role to SNRs in terms of investigations of
the physics of high Mach number shocks and cosmic ray acceleration.
However, they provide access to higher mass densities, radiation
backgrounds, and magnetic field energy densities than found for
supernova remnants, thus enabling studies in previously unexploited
regimes, and due to recent observational detections of non-thermal
emission at hard X-ray and $\gamma$-ray energies are slowly gaining
popularity in the community. Future high energy observational
prospects with Fermi and CTA, etc., look very good. An exciting future
is also expected in the radio regime, with EVLA, e-Merlin, and SKA
(see Benaglia, these proceedings, for further information).

Future theoretical work should combine the best features of the
different modelling efforts to date: for instance, adding a
calculation of the non-thermal particle energy spectra and anisotropic
IC to the hydrodynamical based models of \citet{Pittard:2006b}.  A
more distant goal would be to incorporate the effects of particle
acceleration on the underlying thermal plasma, since DSA, when
present, appears to be very efficient at placing energy into the
non-thermal particle distribution.

\acknowledgements I gratefully acknowledge the invitation of the SOC
to present a review at this conference, and would like to especially
thank Dr. Josep Mart\'{i} for his tireless organization and enthusiasm
which made, I believe, a very successful meeting. I would also like to
thank collaborators and colleagues for their interesting discussions
and input over the years, especially Drs. Sean Dougherty and Don
Ellison, and my PhD student Ross Parkin. Finally, I would like to
thank the Royal Society for funding a Research Fellowship.

\end{document}